\documentclass[twocolumn,showpacs,floats,floatfix,superscriptaddress,aps,pra]{revtex4-1}
\usepackage{amsfonts,amssymb,amsmath}
\usepackage{color,calc}
\usepackage{graphicx}
\usepackage{bmpsize}
\usepackage{bm}
\usepackage[normalem]{ulem}

\def\be{ \begin{equation} }
\def\ee{ \end{equation} }
\def\bea{ \begin{eqnarray} }
\def\eea{ \end{eqnarray} }
\def\bse{ \begin{subequations} }
\def\ese{ \end{subequations} }

\def\i{\,\text{i}}
\def\e{\,\text{e}}
\def\i{i}
\def\e{e}

\def\to{\rightarrow}

\newcommand{\ket}[1]{\vert #1\rangle}
\def\d{\text{d}}

\def\H{\mathbf{H}}
\def\c{\mathbf{c}}

\def\ket#1{| #1 \rangle}
\def\bra#1{\langle #1 |}

\def\to{\rightarrow}
\def\jmax{j_{\text{max}}}
\def\jmax{j_0}
\def\i{\text{i}}

\newcommand{\pryso}{$\text{Pr}^{3+}\text{:}\text{Y}_2\text{SiO}_5\:$}

\begin{document}

\author{Genko T. Genov}
\affiliation{Institut f{\"u}r Angewandte Physik, Technische Universit{\"a}t Darmstadt, Hochschulstr. 6, 64289 Darmstadt, Germany}
\affiliation{Institute for Quantum Optics, Ulm University, Albert-Einstein-Allee 11, 89081 Ulm, Germany}
\author{Marcel Hain}
\email{marcel.hain@physik.tu-darmstadt.de}
\affiliation{Institut f{\"u}r Angewandte Physik, Technische Universit{\"a}t Darmstadt, Hochschulstr. 6, 64289 Darmstadt, Germany}
\author{Nikolay V. Vitanov}
\affiliation{Department of Physics, St. Kliment Ohridski University of Sofia, 5 James Bourchier blvd., 1164 Sofia, Bulgaria}
\author{Thomas Halfmann}
\homepage{http://www.iap.tu-darmstadt.de/nlq}
\affiliation{Institut f{\"u}r Angewandte Physik, Technische Universit{\"a}t Darmstadt, Hochschulstr. 6, 64289 Darmstadt, Germany}

\title{Universal Composite Pulses for Efficient Population Inversion with an Arbitrary Excitation Profile}

\date{January 22, 2020}

\begin{abstract}
We introduce a method to rotate arbitrarily the excitation profile of universal broadband composite pulse sequences for robust high-fidelity population inversion. These pulses compensate deviations in \emph{any} experimental parameter (e.g. pulse amplitude, pulse duration, detuning from resonance, Stark shifts, unwanted frequency chirp, etc.) and are applicable with \emph{any} pulse shape. The rotation allows to achieve higher order robustness to any combination of pulse area and detuning errors at no additional cost.
The latter can be particularly useful, e.g., when detuning errors are due to Stark shifts that are correlated with the power of the applied field.
We demonstrate the efficiency and universality of these composite pulses by experimental implementation for rephasing of atomic coherences in a \pryso crystal.
\end{abstract}

\maketitle


Composite pulses (CPs) have been used for decades in nuclear magnetic resonance \cite{NMR} and since even earlier in applied optics as a tool to design polarization filters and achromatic polarization retarders \cite{Optics}. Recently,
they have also been applied in quantum information processing \cite{Wunderlich,Wang2012NatComm,Ivanov11PRA} and quantum optics for highly accurate and robust qubit rotations \cite{QOptics,Torosov11PRA,Torosov11PRL,SIvanov13NJP}, composite quantum gates \cite{Hill2007,Jones2013pla,Jones2013pra,Merrill2014review,Cohen2016,Calderon-Vargas2017,Tomita2010},
rephasing of atomic coherences \cite{Schraft13PRA,Genov2014PRL,Genov2018PRA}, fault-tolerant dynamical decoupling \cite{RDD_review12Suter,CasanovaPRA2015,GenovPRL2017,SriarunothaiQST2019,GenovQST2019,ZhouArxuv2019},
and robust composite pulses spectroscopy \cite{Dunning2014PRA,Vitanov2015PRA,Zanon-Willette2018REPP}. Example recent applications include also experiments in trapped ions
\cite{Gulde2003,Schmidt-Kaler2003,Haffner2008,Timoney2008,Monz2009,Shappert2013,Mount2015},
 neutral atoms \cite{Rakreungdet2009},
 cold-atom interferometry \cite{Butts2013,Dunning2014,Berg2015},
 optically dense atomic ensembles \cite{Demeter2016},
 quantum dots \cite{Wang2012,Kestner2013,Wang2014,Zhang2017,Hickman2013,Eng2015},
 NV centers in diamond \cite{Rong2015,Aiello2013,SchwartzSciAdv2018}, and optomechanics \cite{Ventura2019}.
The basic idea of CPs is to correct the imperfect interaction of a quantum system with a single pulse by using a sequence of pulses with suitably chosen relative phases. The latter serve as control parameters to choose
an optimized excitation path in Hilbert space, which increases fidelity and robustness with respect to certain
errors.

A common feature of CPs is that they compensate experimental variations in a single parameter only (e.g. pulse duration, pulse amplitude, detuning), or simultaneous fluctuations in at most two parameters \cite{NMR}. Moreover, the optimal phases of CPs usually depend on their pulse shape.
Recently, we derived theoretically and demonstrated experimentally universal CPs for complete population inversion,
which compensate variation in \emph{any} experimental parameter and work with \emph{any} pulse shape \cite{Genov2014PRL}.
The only assumptions made are those of a two-state system, coherent evolution and identical pulses in the CP sequence with accurate control of their relative phases.
These pulses exhibit a remarkably robust performance for any systematic error. We also showed that the obtained solutions for composite pulses consist of two main types of sequences (which we termed ``a'' and ``b'' group), which compensate (with no additional cost) even higher order errors in pulse area and detuning, respectively \cite{Genov2014PRL}.

In this paper, we describe a general theoretical procedure to derive universal composite pulses with an arbitrary rotation of the excitation profile with respect to pulse area and detuning errors. These pulses compensate variation in \emph{any} experimental parameter and work with \emph{any} pulse shape. Additionally, they also allow for higher order error compensation for any combination of pulse area and detuning errors at no additional cost. For example, they could be very efficient for  compensation of Stark shift errors, where the detuning is correlated with the power of the applied field.
As a basic example of relevance to many applications in quantum physics,
we experimentally demonstrate the concept by rephasing of atomic coherences for coherent optical data storage in a \pryso crystal.

\section{Theoretical Description}
%

\begin{table*}[t]
\caption{Phases $\Phi_{l}$ and the corresponding phases $\phi_{k}$ of universal CPs with $n$ pulses (indicated by the number in the label of the CP).
We nullify the coefficients in Eq. \eqref{Eq:Un11_definition} up to order $\jmax=0$ for $n=3$, $\jmax=2$ for $n=5$ to $9$, $\jmax=4$ for $n=13$ and $\jmax=8$ for $n=25$.
Each phase is defined modulo $2\pi$. The phases $\phi_{k}$ are examples for different $\phi_2$ and are determined from $\Phi_{l}$ and Eq. \eqref{Eq:phi_k}.
We show only two examples for U13 and U25 for compactness of presentation.
In order to permit comparison to \cite{Genov2014PRL}, we include also the labels of the "a" and "b" cases for the CPs, which compensate pulse area and detuning errors to a higher order, respectively.
The excitation dynamics remain the same when we simultaneously add a constant shift to all phases $\phi_{k}$.
The change of sign of all phases or the reversal of pulse order reflects the excitation profile (vs. errors in the pulse area and detuning $\Delta$) around the $\Delta=0$ axis (see text).
}
\begin{tabular}{l l l l l} 
\hline 
Group~ & Phases $\Phi_{l},~l=1\dots n-2$ & Sequence~ & Label a/b & ~~~Phases $\phi_{k},~k=1\dots n$ \\ 
\hline 
U3 & $(1)\pi$ &  U3(90$^{\circ}$) & U3a & $(0,1,0)\pi/2$ \\
 & &  U3(0$^{\circ}$) & U3b & $(0,0,1)\pi$ \\
 & &  U3(45$^{\circ}$) & & $(0,1,6)\pi/4$ \\
 & &  U3(135$^{\circ}$) & & $(0,3,2)\pi/4$ \\
U5 & $(2, 3, 2)\pi/3$ & U5(150$^{\circ}$) & U5a & $(0,5,2,5,0)\pi/6$ \\
 & &  U5(330$^{\circ}$) & U5b & $(0, 11, 2,11,0)\pi/6$ \\
 & &  U5(180$^{\circ}$) & & $(0, 3, 2,4,2)\pi/3$ \\
 & &  U5(0$^{\circ}$) & & $(0, 0, 2,1,2)\pi/3$ \\
U7 & $(6,4,5,4,6)\pi/6$ & U7(165$^{\circ}$) & U7a & $(0, 11, 10, 17,10,11,0)\pi/12$ \\
 & &  U7(345$^{\circ}$) & U7b & $(0, 23, 10, 5,10,23,0)\pi/12$ \\
 & &  U7(180$^{\circ}$) & & $(0, 6,6,10,7,8,3)\pi/6$ \\
 & &  U7(0$^{\circ}$) & & $(0, 0,6,4,7,2,3)\pi/6$ \\
U13 & $(12,16,14,16,16,11,16,16,14,16,12)\pi/12$~~~~~~~~~~~& U13(67.5$^{\circ}$) & U13a & $(0, 9, 42, 11, 8, 37, 2,37,8,11,42,9,0)\pi/24$ \\
 & &  U13(247.5$^{\circ}$) & U13b & $(0, 33, 42, 35, 8, 13, 2,13,8,35,42,33,0)\pi/24$ \\
U25 & $(2, 3, 2, 2, 3, 2, 3, 2, 4, 1, 2, 3, 2, 1, 4, 2, 3, 2, 3,$~~~& U25(150$^{\circ}$) & U25a & $(0, 5, 2, 5, 0, 11, 4, 1, 4, 11, 2, 7, 4,7,2,11,4,1,$ \\
& $2, 2, 3, 2)\pi/3$ & & & $4,11,0,5,2,5,0)\pi/6$ \\
 & &  U25(330$^{\circ}$) & U25b & $(0, 11, 2, 11, 0, 5, 4, 7, 4, 5, 2, 1, 4,1,2,5,4,7,$ \\
& & & & $4,5,0,11,2,11,0)\pi/6$ \\ 
\hline 
\end{tabular}
\label{table:coef} 
\end{table*}

We consider a coherently driven two-state quantum system.
Its dynamics obeys the Schr\"{o}dinger equation, $\i \hbar\partial_t \mathbf{c}(t) = \H(t)\mathbf{c}(t)$,
where the vector $\c(t) = [c_1(t), c_2(t)]^T$ contains the probability amplitudes of the two states.
The Hamiltonian in the rotating-wave approximation reads
$\mathbf{H}(t) = (\hbar/2)\Omega(t)\e^{-\i\delta(t)} \ket{1}\bra{2} +\text{H.c.} $,
with
$\delta(t)=\int_{0}^{t}\Delta(t^{\prime})\d t^{\prime}$, where $\Delta=\omega_0-\omega$ is the detuning between the field frequency $\omega$ and the Bohr transition frequency $\omega_0$.
The Rabi frequency $\Omega(t) =-\mathbf{d}\cdot\mathbf{E}(t)/\hbar$ defines the coupling of the two states, induced by the electric field $\mathbf{E}(t)$ and the transition dipole moment $\mathbf{d}$.
In general, both $\Omega(t)$ and $\Delta(t)$ are time-dependent.

Our objective is to achieve complete population inversion in a two-state quantum system even when the properties of the driving pulses are \emph{unknown}. We assume that the composite pulse duration is shorter than the decoherence time of the system, so its evolution due to a single pulse can be characterized by
a propagator
\begin{equation} \label{2stateU}
\mathbf{U}(\alpha,\beta) = \left[\begin{array}{cc}  \epsilon \e^{\i\alpha}  & \sqrt{1-\epsilon^2} \e^{\i\beta} \\  -\sqrt{1-\epsilon^2} \e^{-\i\beta} & \epsilon \e^{-\i\alpha} \end{array}\right],
\end{equation}
where the phases $\alpha$ and $\beta$ and the error term $\epsilon\in[0,1]$ are unknown.
The propagator connects the probability amplitudes at the initial and final times $t_{\text{i}}$ and $t_{\text{f}}$:
$\mathbf{c}(t_{\text{f}})=\mathbf{U}(\alpha,\beta) \mathbf{c}(t_{\i})$.
The transition probability of the single pulse yields $P_{12}^{(1)}=1-|U_{11}|^2= 1-\epsilon^2$.
A constant phase shift $\phi$ in the Rabi frequency leads to the shift $\beta\to\beta+\phi$ in the propagator
$\mathbf{U}(\alpha,\beta+\phi)$.
Then, the propagator of a composite sequence of $n$ identical pulses, each with a phase $\phi_{k}$, reads
\begin{align}
\mathbf{U}^{(n)}\label{Eq:Un_propagator} =&\mathbf{U}(\alpha,\beta+\phi_{n})\dots\mathbf{U}(\alpha,\beta+\phi_{2})\mathbf{U}(\alpha,\beta+\phi_1).
\end{align}
We make no assumptions about the individual pulses in the composite sequence, i.e., how $\epsilon$, $\alpha$ and $\beta$ depend on the interaction parameters.
This justifies the term ``universal'' for these composite pulses because they will compensate imperfections in \emph{any} interaction parameter. We only assume that the constituent pulses are identical and that we can control their phases $\phi_{k}$.
We also note that as we work with the pulse propagator, we make no assumptions about the initial state of the system and population inversion is improved for arbitrary initial states.

In previous work on composite pulses \cite{Genov2014PRL}, it proved useful to apply the so-called anagram condition and choose $\phi_{k}=\phi_{n-k+1}$ in order to obtain the simplest solutions for the phases. In this work, we \emph{do not} make this assumption, which allows us to expand significantly the range of solutions for the universal composite pulses.
First, we define the phase rotation matrix $\mathbf{R}(\phi)\equiv  \exp{\left(-\i \phi \bm{\sigma}_{z}/2\right)}$, so the propagator in Eq. \eqref{2stateU} takes the form
$\mathbf{U}(\alpha,\beta) =  \mathbf{R}^{\dagger}(\beta+\alpha)\mathbf{U}(0,0)\mathbf{R}(\beta-\alpha)$ and the composite pulse propagator in Eq. \eqref{Eq:Un_propagator} becomes
\begin{align}
\mathbf{U}^{(n)}\label{Eq:Un_propagator_expanded}
=&\mathbf{R}^{\dagger}(\phi_n+\beta+\alpha)\mathbf{U}(0,0)\mathbf{R}(\Delta\phi_{n-1}-2\alpha)\mathbf{U}(0,0)\dots\notag\\
 &\mathbf{U}(0,0)\mathbf{R}(\Delta\phi_1-2\alpha)\mathbf{U}(0,0)\mathbf{R}(\phi_1+\beta-\alpha),
\end{align}
where $\Delta\phi_{k}\equiv\phi_{k+1}-\phi_{k},k=1,\dots n-1$. The propagators $\mathbf{R}^{\dagger}(\phi_n+\beta+\alpha)$ and $\mathbf{R}(\phi_1+\beta-\alpha)$ in Eq. \eqref{Eq:Un_propagator_expanded} do not affect the population transfer efficiency of the composite pulse as they cause only phase rotations. Thus, the performance depends only on the phase shifts
$\Delta\phi_{k}-2\alpha$ and the error term $\epsilon$, which affects $\mathbf{U}(0,0)$.

We note that the ``universal'' improvement in performance implies that it should take place for any $\alpha$. Physically, $\alpha$ is typically a phase that is accumulated due to dephasing during (and/or between) the pulses (if they are time-separated). As $\alpha$ is present in all phase shifts $\Delta\phi_{k}-2\alpha$, a universal composite pulse should also improve performance if all $\Delta\phi_{k}$ are shifted by the same phase. Thus, the ``universal'' improvement, i.e., for any phases $\alpha$, $\beta$ and error term $\epsilon$ of the individual pulse, is only due to
the relative phase shifts $\Phi_{k}\equiv\Delta\phi_{k+1}-\Delta\phi_{k}=\phi_{k+2}-2\phi_{k+1}+\phi_{k},~k=1\dots n-2$ (see Appendix for a detailed analysis).
The specific choice of one of the phase shifts, e.g., $\Delta\phi_{1}=\phi_2-\phi_1$, for a given set of $\Phi_{k}$ allows for selective higher order compensation of particular errors, e.g., in pulse area or detuning or combinations of these. In the following, we take $\phi_1=0$ without loss of generality because this assumption amounts to fixing the overall phase of the wave function. Then, we use the phase shift $\Delta\phi_{1}=\phi_2$ as a free parameter to achieve selective higher order error compensation for specific errors. Finally, one can use these assumptions and the definition of $\Phi_{k}$ to express all
phases $\phi_{k}$ of a composite pulse of in terms of $\phi_2$ and $\Phi_{k}$ as
%
%
\begin{equation}\label{Eq:phi_k}
\phi_{k}=(k-1)\phi_2+\sum_{l=1}^{k-1}(k-l-1)\Phi_{l}.
\end{equation}
%
%
%

In order to determine the actual phases of a universal CP sequence of $n$ pulses we calculate the propagator element
\begin{equation}\label{Eq:Un11_definition}
U^{(n)}_{11}=\sum_{j=1}^{n} c_{nj} \epsilon^{j},
\end{equation}
where the coefficients $c_{nj}$ depend on $\alpha$, $\phi_{2}$, and $\Phi_{k}$. Our goal is to nullify (or minimize) all the coefficients $c_{nj}$ up to the highest possible order of $j$, which we label $\jmax$.
As already shown, the proper choice of $\Phi_{k}$ is sufficient to nullify (minimize) the coefficients $c_{nj}$, i.e., achieve ``universal'' improvement in performance.
We first consider the simplest case of a three-pulse sequence ($n=3$) and obtain
\begin{equation}
U_{11}^{(3)}=-(\underbrace{2\e^{\i\xi/2}[\cos{(\Phi_1/2)}] \e^{\i\alpha}+\e^{\i\xi}\e^{-\i\alpha}}_{c_{31}})\epsilon + O(\epsilon^3),
\end{equation}
where $\xi=2\phi_2+\Phi_1$. It is evident that we cannot nullify $c_{31}$ for every $\alpha$ by a proper choice of the control parameters $\Phi_1$ and $\phi_2$. Nevertheless, we can minimize $|c_{31}|=1$ by choosing $\Phi_1=\pi$. Then, the phase $\phi_3=2\phi_2+\pi$ is a general solution for the universal composite pulse for population inversion of three pulses. The free parameter $\phi_2$ can be used to achieve higher order error compensation for specific errors. We note that by choosing $\phi_2=\pi/2$, we obtain $\phi_3=0$, i.e., the solution for the U3 composite pulse in \cite{Genov2014PRL}.

%
Next, we give an example for a five-pulse sequence ($n=5$), where we obtain
\begin{align}
U_{11}^{(5)} =&\underbrace{\e^{\i(2\phi_2+\Phi_1+\Phi_2)}\left(c_{51+}\e^{\i\alpha}+\e^{\i(\phi_2+\Phi_1+\Phi_3)}c_{51-}\e^{-\i\alpha}\right)}_{c_{51}}\epsilon \notag \\
&+ O(\epsilon^3),
\end{align}
where
\begin{align}
c_{51+} &= 1 + \e^{\i\Phi_3}\left(1+\e^{\i\Phi_1}\right),\\
c_{51-} &= 1+\e^{\i\Phi_2}.\notag
\end{align}
Then, we can nullify $c_{51}$ for any $\alpha$ by nullifying both $c_{51+}$ and $c_{51-}$, which requires the anagram condition $\Phi_1=\Phi_3$, and choosing $\Phi_1=2\pi/3$ and $\Phi_2=\pi$.
As the error term $\epsilon$ is typically small, the composite pulse transition probability $P_{12}^{(5)} = 1 - O(\epsilon^6)$ is much closer to unity than the transition probability of a single pulse $P_{12}^{(1)} = 1 -\epsilon^2$.
The phases of the individual pulses in the sequence $\phi_k$ can be determined by the formula in Eq. \eqref{Eq:phi_k} and take the form $\phi_3=2\phi_2+2\pi/3$, $\phi_4=3\phi_2+\pi/3$, $\phi_5=4\phi_2+2\pi/3$. This is a general solution for the universal composite pulse for population inversion, consisting of five pulses. The free parameter $\phi_2$ can be used to achieve higher order error compensation for specific errors. We note that by choosing $\phi_2=5\pi/6$ and $\phi_2=11\pi/6$, we obtain the solutions for the U5a and U5b composite pulses in \cite{Genov2014PRL}.

We can derive the phases of universal composite pulses with a higher number of pulses in an analogous way. Specifically, we nullify the coefficients $c_{nj}$ up to the maximum possible order $\jmax$ and minimize the absolute values of the coefficients of the first non-zero order $\jmax+1$.
The transition probability for such CP reads
$P_{12}^{(n)} = 1-|U^{(n)}_{11}|^2 = 1-O(\epsilon^{2\jmax+2})$.
%
As the error term $\epsilon$ is typically small, the composite pulse transition probability $P^{(n)}$ quickly converges towards unity by using longer composite sequences. The length of pulse sequences is only limited by the coherence time of the quantum system, to guarantee that the effects of all pulses in the sequence remain coherent to each other. 
In order to incorporate the continuous parameter $\phi_2$ in the name of the universal CP sequences we name them ``U$n$($\phi_2$)'' where $n$ is the number of pulses in the sequence. For instance, the U5a and U5b sequences in \cite{Genov2014PRL} are named U5(150$^{\circ}$) and U5(330$^{\circ}$) in this new naming scheme.

%

Several universal CPs are listed in Table \ref{table:coef}. The phases $\Phi_{l}$ are the same and unique (up to the sign of all of them) for the CP with a particular number of pulses. The proper choice of the free parameter $\phi_2$ allows to achieve higher order error compensation for specific errors or their combinations. For example, the CPs labeled {``a''} perform better against variations in the pulse duration, whereas those labeled {``b''} perform better against the detuning \cite{Genov2014PRL}, while the other sequences allow for additional error compensations when the errors are correlated. The latter can be especially useful, e.g., when the detuning errors are due to Stark shifts that are correlated to the power of the driving field.
All of these, however, permit compensation against both parameters, as well as against \emph{all} other parameters.
%

\begin{figure}[t!]
\includegraphics[width=\columnwidth]{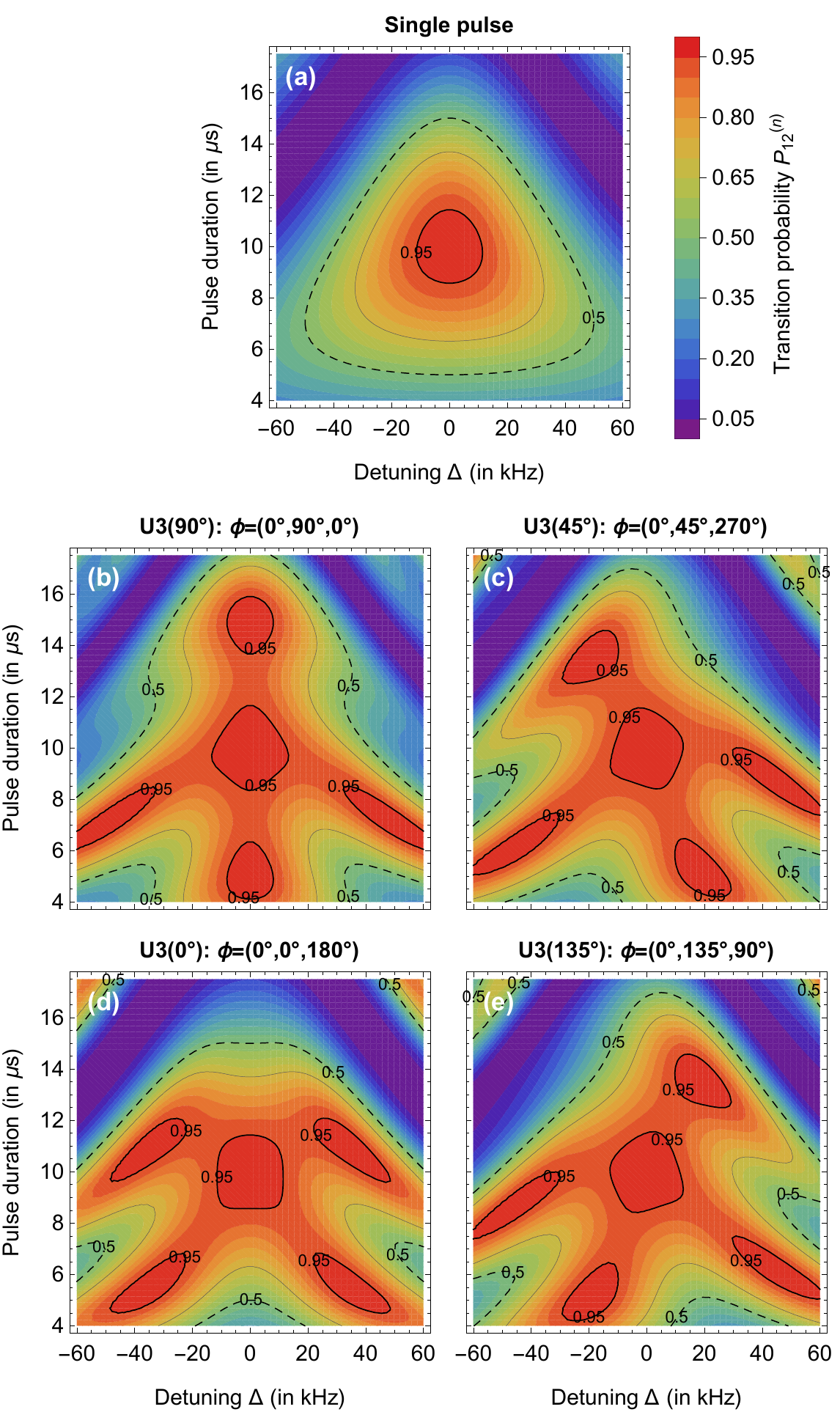}
\caption{(color online)
Numerical simulation: Transition probability $P^{(n)}_{12}$ vs. static detuning and duration $T$ of each constituent pulse (referred to as ``pulse duration'') for
a single pulse (a) and CPs from the U3 group (b - e) with phases from Table \ref{table:coef}. The particular choice of $\phi_2$ allows arbitrary rotation of the excitation profile and higher order error compensation at no additional cost. Note that the excitation profiles of U3(90$^{\circ}$) and U3(0$^{\circ}$) are symmetric around the $\Delta=0$ axis. The U3(90$^{\circ}$) [U3(0$^{\circ}$)] pulses compensate higher order errors with respect to pulse area [detuning] and are thus also labeled U3a [U3b] in Table \ref{table:coef}. The pulses are assumed rectangular with a Rabi frequency of $\Omega=50$kHz (in frequency units). Thick solid lines indicate regions of high transfer probabilities beyond 0.95, thin solid lines indicate regions with transfer probability beyond 0.7 and dashed lines indicate regions with transfer probability beyond 0.5.
 }
\label{Fig-cp3}
\end{figure}

\begin{figure*}[t!]
\includegraphics[width=\textwidth]{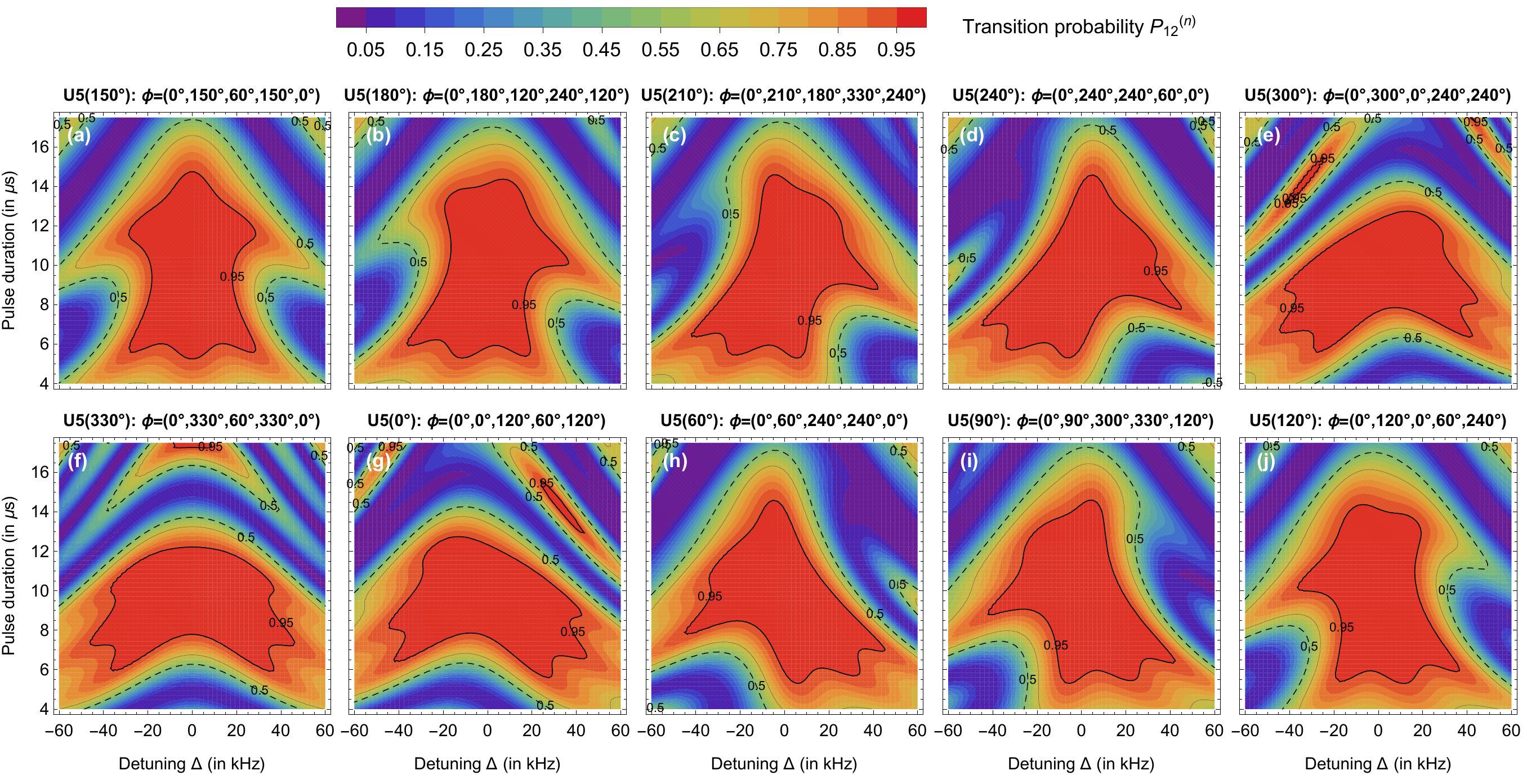}
\caption{(color online)
Numerical simulation: Transition probability $P^{(n)}_{12}$ vs. static detuning and duration $T$ of each constituent pulse for
CPs from the U5 group with phases from Table \ref{table:coef} and Eq. \eqref{Eq:phi_k}.
As expected, the high efficiency ranges are much broader than with the U3 sequence [compare with Fig. \ref{Fig-cp3} (b - e)].
As we change the free parameter $\phi_2$, the excitation profile rotates and allows additional higher order error compensation for specific pulse area and detuning errors, as well as their combinations. Note that the excitation profiles of U5(150$^{\circ}$) and U5(330$^{\circ}$) are symmetric around the $\Delta=0$ axis. The pulses are assumed rectangular with a Rabi frequency of $\Omega=50$~kHz (in frequency units). For compactness of presentation we omit the plots for U5(270$^{\circ}$) and U5(30$^{\circ}$).
 }
\label{Fig-cp5}
\end{figure*}

The performance of the universal composite pulses for different phases $\phi_2$ versus deviations in the pulse duration $T$ of each constituent pulse and the detuning is shown in figures~\ref{Fig-cp3} and~\ref{Fig-cp5}.
Figure~\ref{Fig-cp3} compares a single pulse (a) to four different three-pulse CPs from the U3 group (b - e). As it is well known, the transition probability for a single pulse [Fig.~\ref{Fig-cp3} (a)] quickly drops when the pulse duration $T$ does not match to a perfect $\pi$-pulse, i.e., $T=10 \mu$s, or when the pulse carrier frequency is off resonance.
All U3 universal CPs are robust with respect to variations along both pulse area and detuning errors. Hence, the areas of high transfer probability (e.g. beyond 0.95 or 0.5, see thick solid and dashed lines in Fig.~\ref{Fig-cp3}) increase for excitation with composite pulses compared to a single pulse.
The U3(90$^{\circ}$) sequence compensates higher order errors with respect to pulse duration, i.e., pulse area, and is thus labeled also U3a. Similarly, U3(0$^{\circ}$) corrects higher order detuning errors and is labeled U3b. 
Moreover, by variation of $\phi_2$, we can achieve higher order compensation for specific combinations of these errors. This is clearly visible in the simulations by a rotation and controlled distortion of the regions of high transfer efficiency, when the phase $\phi_2$ varies [compare the shape of the patterns in Fig.~\ref{Fig-cp3} (b - e)]. Specifically, the U3(45$^{\circ}$) [U3(135$^{\circ}$)] CP compensate higher order errors when the Rabi frequency and detuning errors have a negative [positive] correlation. The latter can be particularly useful when detuning errors are due to Stark shifts, which are correlated with the power of the applied field.
 %

We note that U3 is not a truly universal CP because the phases do not nullify the first-order coefficients (i.e., $\jmax=0$), but only minimize them, cf. Table~\ref{table:coef}.
Figure~\ref{Fig-cp5} shows the performance of the higher-order, genuine universal U5 pulses for different $\phi_2$. Again, the U5(150$^{\circ}$) and U5(330$^{\circ}$) CPs compensate higher order errors with respect to the pulse duration or detuning, respectively \cite{Genov2014PRL}. Also for the five-pulse sequences, variation of $\phi_2$ rotates and controllably distorts the composite efficiency patterns to enable higher order compensation for specific combinations of pulse errors. The optimum choice of the CP will depend on the particular errors and their correlation. For example, the U5(60$^{\circ}$) [U5(240$^{\circ}$)] CP have a higher order error compensation when the Rabi frequency and detuning errors have a negative [positive] correlation.

We note that for all universal CPs the change of the sign of the phases $\phi_k$ or the reversal of pulse order reflects the excitation profile with respect to the $\Delta=0$ axis. The same reflection takes place when the phases $\phi_{k}$ are calculated from Eq. \eqref{Eq:phi_k} with the values of $\Phi_{l}$ in Table \ref{table:coef}, $\phi_2=(\phi_2^{a,b}+\widetilde{\phi})$ and we change the sign of $\widetilde{\phi}$. In the latter expression, $\phi_2^{a,b}$ are the phases $\phi_2$ of the U$n$a and U$n$b CPs, respectively (see Table \ref{table:coef}), and $\widetilde{\phi}$ is arbitrary.

We verified by extensive simulations the robustness of the universal CPs of seven and more pulses against variations in other interaction parameters, e.g. Stark shifts, unwanted frequency chirp, pulse shape, frequency jitter, etc.
All simulations confirm that our universal CPs are amazingly robust to any such variation and the high-fidelity region expands steadily with the CP order. 
We note that the universal composite pulses can improve performance even when pulse parameters, e.g., amplitude or frequency, vary on a time scale, which is shorter than individual pulse duration as long as it is systematic, i.e., it is repeated in every pulse. 
Again, for all universal CPs, shifting $\phi_2$ allows us to achieve higher order compensation for specific combinations of these errors.

\section{Experimental Demonstration}

We experimentally verified the performance of 
universal CP sequences by rephasing atomic coherences for optical data storage.
In the experiment, we generate the atomic coherence on a radio-frequency (RF) transition
between two inhomogeneously broadened hyperfine levels of a \pryso crystal.
The atomic coherence between the two quantum states is optically prepared and read-out by electromagnetically-induced transparency (EIT) \cite{Fleischhauer05RMP}, which enables a straightforward optical readout.
The EIT scheme couples states $\vert 1\rangle$ and $\vert 2 \rangle$ by a strong control field and a weak probe field via an excited state $\vert 3\rangle$.
By simultaneously and adiabatically turning off the control and probe fields, we convert the probe field into an atomic coherence, i.e. a coherent superposition of states $\vert 1\rangle$ and $\vert 2 \rangle$. This is the ``write'' process of optical information encoded in the probe field, often termed ``stopped light'' or ``stored light'' \cite{Fleischhauer05RMP,Heinze13PRL,Schraft16PRL}.
To ``read'' the optical memory after an arbitrary storage time, we apply the strong control field again to beat with the atomic coherence and thereby generate a signal field with the same properties as the stored field.
The concept and the experimental setup for EIT-based light storage in \pryso are described in detail elsewhere
\cite{Schraft16PRL,Mieth14OE}. 
After optimizing the parameters of the light storage process we achieve an efficiency of $(20.0\pm0.4)\%$ for a readout after $2\,\mu$s which is much shorter than the dephasing time of about $20\,\mu$s.

In such a coherent optical memory it is crucial to reverse the effect of dephasing of the atomic coherences during the storage time.
The dephasing is due to inhomogeneous broadenings of the hyperfine levels.
Rephasing is implemented usually by resonant RF $\pi$-pulses, e.g. in a standard Carr-Purcell-Meiboom-Gill (CPMG) sequence \cite{CPMG}.
However, resonant $\pi$-pulses do not work efficiently in systems with large inhomogeneous broadening, as the transition frequency varies for different ions.
The efficiency is further reduced by the spatial inhomogeneity of the RF field over the crystal.
As an alternative, adiabatic rephasing techniques, e.g. rapid adiabatic passage \cite{Pascual-Winter12PRB,Mieth12PRA} or composite adiabatic passage \cite{Torosov11PRL,Schraft13PRA} offer improved operation bandwidth.
To permit much broader operation bandwidth, we replace now the single $\pi$-pulses in the CPMG rephasing sequence by our universal CPs. In the experiment we set the storage time to $400\,\mu$s, i.e. much larger than the dephasing time.
The optical ``write'' and ``read'' sequences were kept fixed, while the RF rephasing pulses were varied.
Therefore, the energy of the retrieved optical signal 
serves as a measure for the rephasing efficiency, and hence, the efficiency of the driving $\pi$-pulse or CP.

The different rephasing pulses are generated by an arbitrary waveform generator (AWG 5014, Tektronix) and amplified to a maximum power of $30$~W (LZY-22+, Mini-Circuits). In order to maximize the power of the RF pulses emitted by our pair of coils surrounding the \pryso crystal we use a single frequency impedance matching circuit. This circuit provides a 3~dB bandwidth of about 600~kHz which is much wider than the maximum detuning range of 120~kHz considered in this paper.

\begin{figure}
\includegraphics[width=\columnwidth]{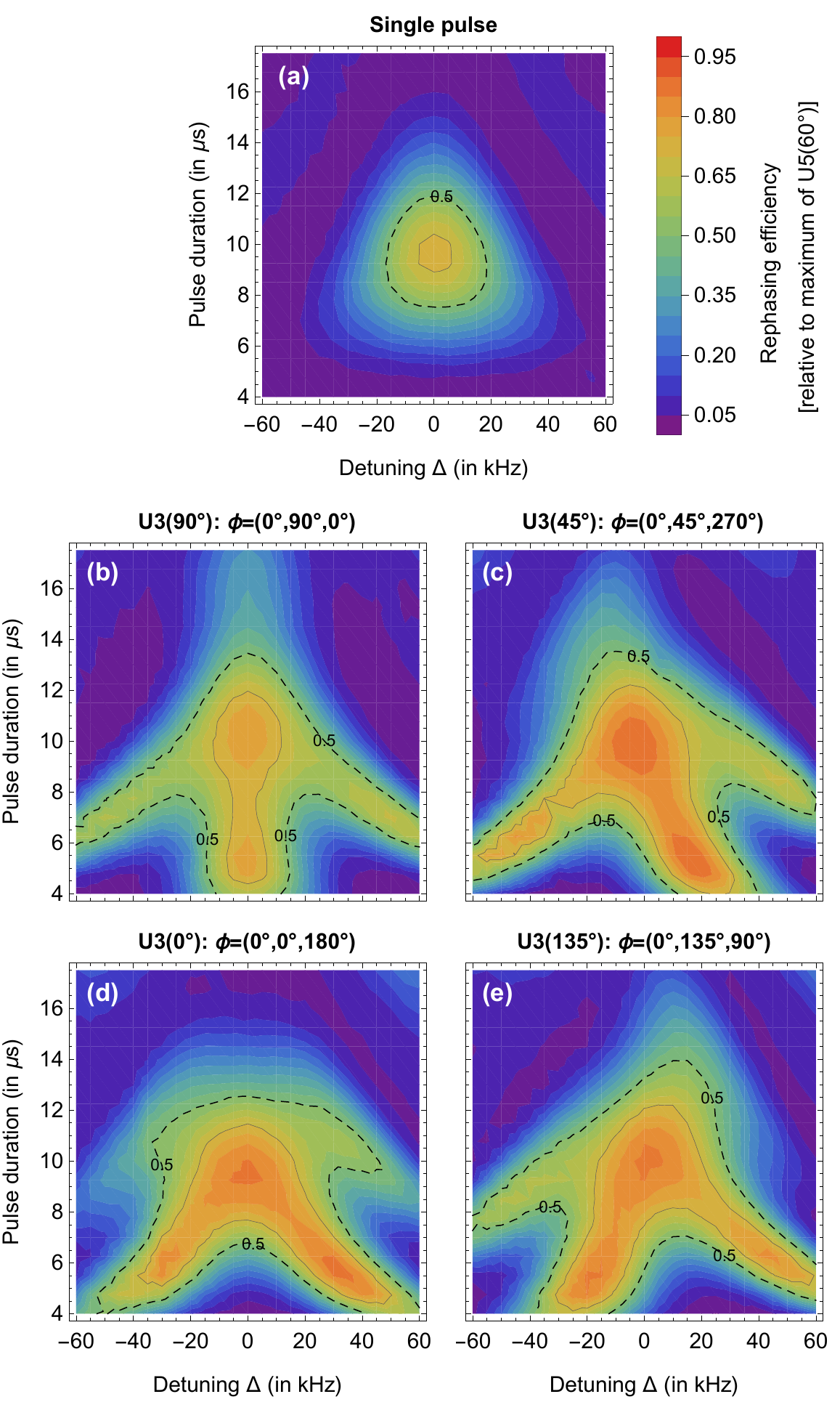}
\caption{(color online)
Experimental data: Rephasing efficiency in the EIT-driven optical memory in \pryso, involving either single pulses (a) or different U3 rephasing sequences (b - e), vs. pulse duration and detuning.
Blue color indicates low, red color indicates high rephasing efficiency. The rephasing efficiency is normalized with respect to the maximum efficiency of all investigated sequences [i.e., the maximum of the U5(60$^{\circ}$) sequence].
 }
\label{U3_exp}
\end{figure}

We set the amplitude of the RF pulses such that the Rabi frequency is constant
and perform a systematic measurement of the rephasing efficiency for different detuning and duration of the pulses.
Figure~\ref{U3_exp} (a) shows the rephasing efficiency of a single pulse relative to the maximum rephasing efficiency of the U5(60$^{\circ}$) sequence,
which performed best in our experiment, as demonstrated later in the text.
The contour plot shows a clear maximum in the rephasing efficiency for zero detuning and a pulse duration of $10~\mu$s.
As the maximum is expected when the applied pulse is a resonant $\pi$ pulse, we estimate that our Rabi frequency is approximately $50$~kHz (in frequency units).
Despite the rephasing the maximum light storage efficiency is reduced to 4.7~\%. The reduction of the efficiency by a factor of four compared to the short storage time of 2~$\mu$s is partially due to pulse errors and partially due to additional decoherence effects which lead to a decoherence time of about 500~$\mu$s, i.e. in the range of our storage time. For pulse parameters deviating from the optimum the transfer efficiency of the $\pi$-pulses, and hence the rephasing efficiency of the light storage experiment, quickly drops.
We note, that for application of two $\pi$-pulses the rephasing efficiency can be approximated by the square of the single-pulse transition probability \cite{Genov2018PRA}.
This leads to a narrowing of the efficient parameter region [compare Fig.~\ref{Fig-cp3} (a)], while the shape of the excitation profile is preserved.

Now we replace each of the single
$\pi$-pulses by three consecutive
$\pi$-pulses with relative phases from the U3 group and repeat the parameter scan. Figure~\ref{U3_exp} (b - e) shows the experimentally determined relative rephasing efficiency vs. detuning and pulse duration of our composite pulse sequences. The shapes of the obtained efficiency patterns clearly confirm the theoretical prediction (compare Fig.~\ref{Fig-cp3}). The U3 composite pulse sequences significantly increase the regions of high rephasing efficiency. Choice of $\phi_2$ rotates the excitation pattern. While the U3(90$^{\circ})$ sequence works well to compensate pulse duration errors at zero detuning, the U3(45$^{\circ}$) and U3(135$^{\circ}$) sequences enable compensation of specific combinations of the two parameters. Hence, when pulse area and detuning errors are correlated in an experiment, we can choose $\phi_2$ in the rephasing composite sequence to cope with any specific correlation and maintain high transfer efficiency. We note, that in figure~\ref{U3_exp} we show only four specific choices of $\phi_2$. Nevertheless, we confirmed that the excitation pattern can be varied continuously with $\phi_2$. This new feature of composite sequences simply requires appropriate choice of phases, while no changes to the experimental setup are necessary.

We note, that careful comparison of our experimental data to the simulated transfer efficiency (see Fig.~\ref{Fig-cp3}) shows that the expected additional smaller areas of higher efficiency in the ``wings'' of the U3(90$^{\circ}$) profile [e.g., at $\Delta=\pm (30...60)$~kHz and pulse durations of $5...9~\mu$s in Fig.~\ref{Fig-cp3} (b)] have less efficiency than theoretically expected. We attribute this to RF field inhomogeneities, which lead to averaging effects and washing out of smaller features in the excitation patterns. Moreover, the experimental data indicate a drop of the rephasing efficiency for larger pulse durations - while the simulation for the U3(90$^{\circ}$ pulse sequence shows a region of high transfer efficiency for zero detuning and a pulse duration of $15~\mu$s. This could be due to inhomogeneous broadening and field inhomogeneities in our experimental setup, which reduce the rephasing efficiency of single pulses for longer interaction times and are not taken into account in our simplified simulations.

\begin{figure}
\includegraphics[width=\columnwidth]{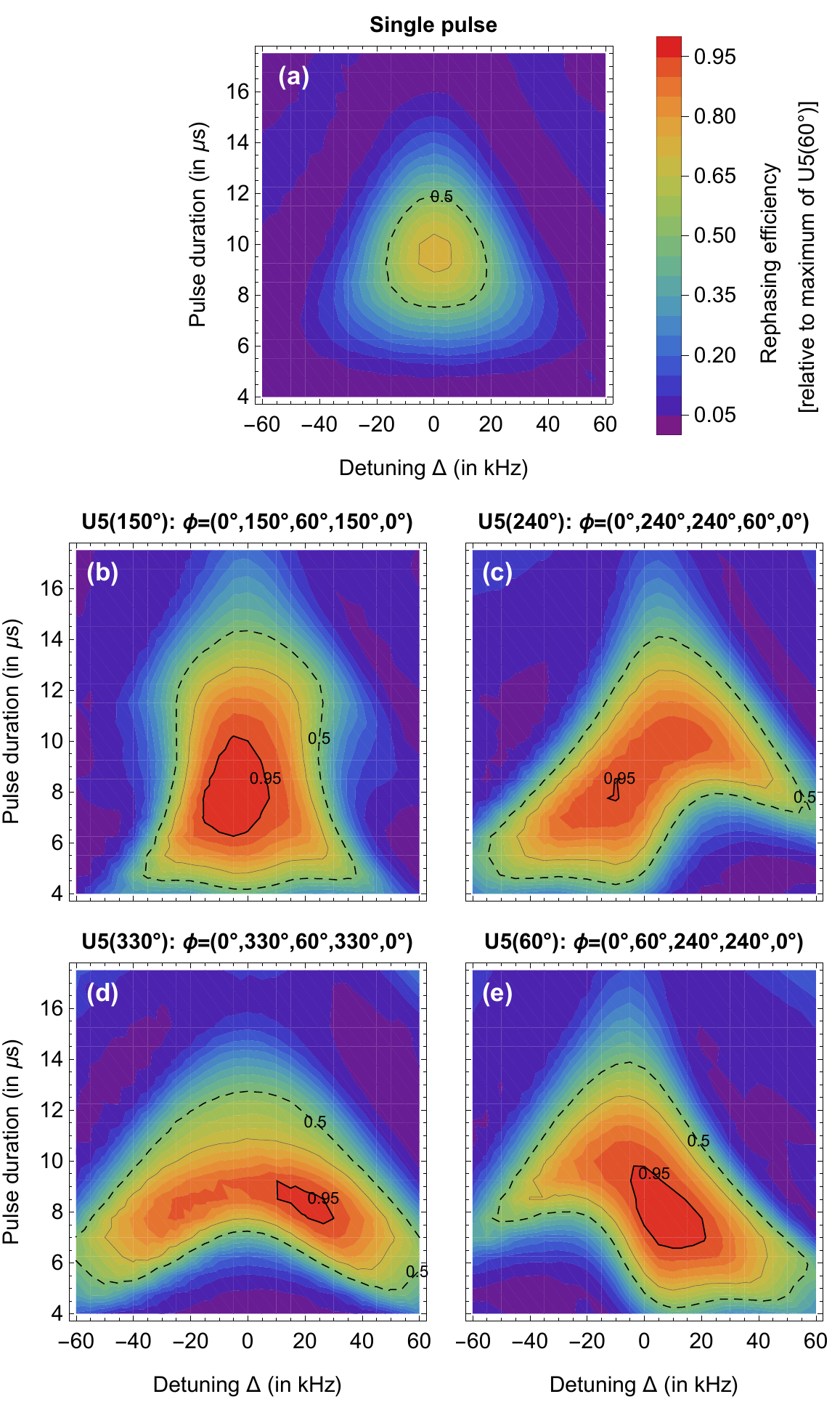}
\caption{(color online)
Experimental data: Rephasing efficiency in the EIT-driven optical memory in \pryso, involving either single pulses (a) or different U5 rephasing sequences (b - e), vs. pulse duration and detuning. The rephasing efficiency is normalized with respect to the maximum efficiency of all investigated sequences [i.e., the maximum of the U5(60$^{\circ}$) sequence].
}
\label{U5_exp}
\end{figure}

In order to proceed towards longer composite sequences in the experiment, we replaced now the U3 pulse sequence by five pulses with phases according to the U5 group. Figure~\ref{U5_exp} shows the parameter scans, similar to figure~\ref{U3_exp}. Very obviously, the rephasing efficiency is now substantially improved, compared not only to the single $\pi$-pulse, but also compared to the U3 sequences. Both the plateaus of efficient transfer efficiency, as well as the peak values increase. The U5(60$^{\circ}$) sequence reaches the maximum rephasing efficiency of all sequences shown in this publication. At a detuning of 5~kHz and a pulse duration of 7.75~$\mu$s the rephasing efficiency is
a factor of 1.37 higher compared
to the simple $\pi$-pulse rephasing. Note that the increase in rephasing efficiency depends on the amount and type of pulse errors and the given value is specific to our experimental setup. Our data confirm the theoretically predicted higher capability for error compensation. Again, comparison of the four U5 sequences shows controlled rotation and distortion of the excitation pattern by variation of phases.

%

\section{Conclusion}
In conclusion, we theoretically developed and experimentally demonstrated generalized universal broadband composite pulses for robust high-fidelity population inversion with the possibility to arbitrarily rotate their excitation profile in parameter space. The relative phases of the pulses in the sequence serve as control parameters to steer the quantum system on robust pathways through Hilbert space.
The pulse sequences compensate deviations in \emph{any} experimental parameter (e.g. pulse amplitude, pulse duration, detuning from resonance, Stark shifts, unwanted frequency chirp, etc.) and are applicable with \emph{any} pulse shape. As the only constraints we require identical pulses in the sequence, accurate control of their relative phases, and the total duration of the pulse sequence must not exceed the coherence time of the medium. The rotation of the excitation profile allows to achieve higher order robustness to any combination of pulse area and detuning errors at no additional cost.

We experimentally demonstrated the concept by systematic studies of the rephasing efficiency of atomic coherences during EIT-based light storage in a \pryso crystal. The experimental implementations of our new U3 and U5 composite pulse sequences agree very well with the numerical simulations and, hence, fully support the theoretical predictions. Both the peak transfer efficiency, as well as the robustness (e.g., measured by the extension of regions of high transfer efficiency) are much larger for rephasing by composite pulses compared to simple $\pi$-pulses. Moreover, the experimental data demonstrate the possibility to rotate the excitation patterns in parameter space by variation of the pulse phases. This is of relevance to any experiment, which requires efficient and robust compensation of correlated excitation pulse errors.
Specific examples include e.g., quantum gates, dynamical decoupling, or composite pulses spectroscopy, where the individual population inversion pulses can be replaced by composite pulses or embedded in the protocol.

\begin{acknowledgments}
GG acknowledges support of the European Union under grant agreement No. 667192-HYPERDIAMOND under the Horizon 2020 program. NVV acknowledges support by the European Commission’s Horizon-2020 Flagship
on Quantum Technologies project 820314 (MicroQC).
\end{acknowledgments}

\appendix*

\section{Proof of improved performance for any $\phi_2$}\label{Appendix}

In this section we show that the ``universal'' improvement in performance for a ``universal'' composite pulse is due only to the choice of the phases $\Phi_{k}$, as defined in the main text, and can be achieved for any phase $\phi_2$.  

First, the propagator of a single pulse is conveniently parameterized by \cite{Genov2014PRL,GenovPRL2017,Genov2018PRA}
\begin{equation}
\mathbf{U}(\alpha,\beta) = \left[\begin{array}{cc} \epsilon \e^{\i\alpha}  & \sqrt{1-\epsilon^2} \e^{\i\beta} \\  -\sqrt{1-\epsilon^2} \e^{-\i\beta} & \epsilon \e^{-\i\alpha} \end{array} \right],
\end{equation}
where the phases $\alpha$ and $\beta$ and an error term $\epsilon\in[0,1]$ are unknown.
A constant phase shift $\phi$ in the Rabi frequency transforms the phase $\beta$ to $\beta+\phi$ and the propagator $\mathbf{U}(\alpha,\beta)$ to
\begin{align}\label{Eq:U_phi_propagator}
\mathbf{U}(\alpha,\beta+\phi) &=  \mathbf{R}^{\dagger}(\phi)\mathbf{U}(\alpha,\beta)\mathbf{R}(\phi) \\
& = \left[\begin{array}{cc} \epsilon \e^{\i\alpha}  & \sqrt{1-\epsilon^2} \e^{\i(\beta+\phi)} \\  -\sqrt{1-\epsilon^2} \e^{-\i(\beta+\phi)} & \epsilon \e^{-\i\alpha} \end{array} \right],\notag
\end{align}
where phase rotation matrix is defined as
\begin{align}
\mathbf{R}(\phi) &\equiv  \exp{\left(-\i \phi \bm{\sigma}_{z}/2\right)} = \left[\begin{array}{cc} \e^{-\i\phi/2}  & 0 \\ 0 & \e^{\i\phi/2} \end{array} \right],\notag
\end{align}
where $\bm{\sigma}_{z}$ is the respective Pauli matrix. We note that the unknown phases $\alpha$ and $\beta$ can also be included in the phase rotation matrix, so the propagator in Eq. \eqref{Eq:U_phi_propagator} takes the form
\begin{align}\label{Eq:U_phi_propagator_V2}
\mathbf{U}(\alpha,\beta+\phi) &=  \mathbf{R}^{\dagger}(\phi+\beta+\alpha)\mathbf{U}(0,0)\mathbf{R}(\phi+\beta-\alpha).
\end{align}

Assuming coherent evolution during a sequence of $n$ pulses with different relative phases $\phi_{k}$, the propagator of the composite sequence then becomes
\begin{equation}\label{Eq:U_phased_sequence}
\mathbf{U}^{(n)}=\mathbf{U}(\alpha,\beta+\phi_{n})\dots \mathbf{U}(\alpha,\beta+\phi_1),
\end{equation}
and the phases $\phi_{k}$ of the individual pulses can be used as control parameters to achieve a robust performance. For example, the propagator of a sequence of $n=3$ pulses with phases $\phi_1$, $\phi_2$ and $\phi_3$ is given by
\begin{align}
\mathbf{U}^{(3)}\label{Eq:U3_propagator} =&\mathbf{U}(\alpha,\beta+\phi_{3})\mathbf{U}(\alpha,\beta+\phi_{2})\mathbf{U}(\alpha,\beta+\phi_1)\notag\\
=&\mathbf{R}^{\dagger}(\phi_3+\beta+\alpha)\mathbf{U}(0,0)\mathbf{R}(\Delta\phi_2-2\alpha)\notag\\
 &\mathbf{U}(0,0)\mathbf{R}(\Delta\phi_1-2\alpha)\mathbf{U}(0,0)\mathbf{R}(\phi_1+\beta-\alpha),
\end{align}
where $\Delta\phi_{k}\equiv\phi_{k+1}-\phi_{k}$ is the relative phase shift between the $(k+1)$-th and the $k$-th pulses.

The propagators $\mathbf{R}^{\dagger}(\phi_3+\beta+\alpha)$ and $\mathbf{R}(\phi_1+\beta-\alpha)$ in Eq. \eqref{Eq:U3_propagator} do not affect the population transfer efficiency as they cause only phase rotations around the $z$ axis of the Bloch sphere. Thus, the population transfer efficiency of the composite pulse can be determined solely from the modified propagator
\begin{align}
\mathbf{\widetilde{U}}^{(3)}=&\mathbf{U}(0,0)\mathbf{R}(\Delta\phi_2-2\alpha)
\mathbf{U}(0,0)\mathbf{R}(\Delta\phi_1-2\alpha)\mathbf{U}(0,0),
\end{align}
which does not depend on the phase $\beta$ but only on the relative phase shifts $\Delta\phi_{k}$ between the pulses, the unknown phase $\alpha$ (e.g., due to dephasing during a pulse), and the error term $\epsilon\in[0,1]$, which quantifies the single pulse population transfer efficiency error: $\epsilon^2\equiv 1-P_{12}^{(1)}$.

The analysis can be generalized for a sequence of $n$ pulses, where the modified propagator takes the form
%

\begin{align}
\mathbf{\widetilde{U}}^{(n)}=&\mathbf{U}(0,0)\mathbf{R}(\Delta\phi_{n-1}-2\alpha)\dots\\
&\mathbf{U}(0,0)\mathbf{R}(\Delta\phi_2-2\alpha)\mathbf{U}(0,0)\mathbf{R}(\Delta\phi_1-2\alpha)\mathbf{U}(0,0).\notag
\end{align}
%
We can redefine all relative phase shifts by taking $\chi_{k}\equiv\Delta\phi_{k+1}-\Delta\phi_{1}$, $k=1,\dots,n-2$ and obtain
\begin{align}
\mathbf{\widetilde{U}}^{(n)}=&\mathbf{U}(0,0)\mathbf{R}(\chi_{n-2}+\widetilde{\alpha})\dots\\
&\mathbf{U}(0,0)\mathbf{R}(\chi_{1}+\widetilde{\alpha})\mathbf{U}(0,0)\mathbf{R}(\widetilde{\alpha})\mathbf{U}(0,0),\notag
\end{align}
where $\widetilde{\alpha}\equiv \Delta\phi_{1}-2\alpha$. 
We note that we can take $\phi_1=0$ without loss of generality as usually only the relative phases between the pulses have a physical meaning. Thus, $\Delta\phi_1=\phi_2-\phi_1=\phi_2$ and $\widetilde{\alpha}= \phi_{2}-2\alpha$.

We note that in our analysis we make no assumptions about the individual pulses in the composite sequence, i.e., how $\epsilon$, $\alpha$ and $\beta$ depend on the interaction parameters.
This justifies the term ``universal'' for our composite pulses because they compensate imperfections in \emph{any} interaction parameter. As a result, they improve performance for any $\alpha$ and thus \emph{for any parameter $\widetilde{\alpha}=\phi_{2}-2\alpha$}. Since $\widetilde{\alpha}$ is a linear combination of $\alpha$ and $\phi_2$, the effect of a change in $\alpha$ by $\Delta\alpha$ is equivalent to a change in $\phi_2$ by $-2\Delta\alpha$ as long as the phases $\chi_{k}$ remain the same. Thus, the requirement that a ``universal'' composite pulse improves performance for any $\alpha$ is equivalent to the requirement that it improves performance for any $\Delta\phi_1=\phi_2$. 
In other words, the ``universal'' improvement in performance for a composite pulse is due only to the choice of the phases $\chi_{k}$, where $k = 1, \dots, n-2 $ as it is achieved for any $\widetilde{\alpha}$ and thus for any phase $\phi_2$.  Moreover, the ``universal'' improvement in performance for any $\phi_2$, given a set of $\chi_{k}$, is a necessary condition for improved performance for any $\alpha$, and thus for a ``universal'' composite pulse.

Finally, it proves useful to define the phases
\begin{align}
\Phi_{k}=\chi_{k}-\chi_{k-1}=\phi_{k+2}-2\phi_{k+1}+\phi_{k},
\end{align}
where $k=2,\dots, n-2$ and $\Phi_1=\chi_1$. These are linear combinations of $\chi_{k}$ and the latter can be expressed as $\chi_{k}=\sum_{i=1}^{k}\Phi_{i}$. The advantage of $\Phi_{k}$ is that the anagram condition $\Phi_{k}=\Phi_{n-k-1},k=1,\dots,n-2$ is required for the nullification or minimization of coefficients $c_{nj}$, as defined in Eq. \eqref{Eq:Un11_definition}, and allows for simplification of the analysis and presentation. Thus, we use them in the main text.

In summary, we showed that the ``universal'' improvement in performance for a composite pulse is due only to the choice of the phases $\chi_{k}$ (and their linear combinations $\Phi_{k}$) and can be achieved for any phase $\phi_2$. The latter is a necessary condition for ``universal'' improvement in performance for any $\alpha$ and thus for a ``universal'' composite pulse. The proper choice of $\phi_2$ allows only for additional higher-order compensation of specific errors, e.g., better compensation of detuning, pulse duration errors, etc. or combinations of these, as we show in the main text.



\begin{thebibliography}{99}

\bibitem{NMR} M.~H.~Levitt, Prog. NMR Spectrosc. \textbf{18}, 61 (1986);
 R.~Freeman, \emph{Spin Choreography} (Spektrum, Oxford, 1997).




\bibitem{Optics} C.~D.~West and A.~S.~Makas, 
 J. Opt. Soc. Am. \textbf{39}, 791 (1949);
M.~G.~Destriau and J.~Prouteau, 
 J. Phys. Radium \textbf{10}, 53 (1949);
S.~Pancharatnam, 
 Proc. Ind. Acad. Sci. \textbf{51}, 130 (1955);
 \emph{ibid.} \textbf{51}, 137 (1955);
S.~E.~Harris, E.~O.~Ammann, and A.~C.~Chang, 
 J. Opt. Soc. Am \textbf{54}, 1267 (1964);
C.~M.~McIntyre and S.~E.~Harris, 
 \textit{ibid.} \textbf{58}, 1575 (1968).


\bibitem{Wunderlich}
M.~Pons, V.~Ahufinger, C.~Wunderlich, A.~Sanpera, S.~Braungardt, A.~Sen(De), U.~Sen, and M.~Lewenstein, Phys. Rev. Lett. \textbf{98}, 023003 (2007);
N.~Timoney, V.~Elman, S.~Glaser, C.~Weiss, M.~Johanning, W.~Neuhauser, and C.~Wunderlich, Phys. Rev. A \textbf{77}, 052334 (2008);
C.~Piltz, B.~Scharfenberger, A.~Khromova, A.~F.~Varon, and C.~Wunderlich, Phys. Rev. Lett. \textbf{110}, 200501 (2013).

\bibitem{Wang2012NatComm}X.~Wang, L.~S.~Bishop, J.~P.~Kestner, E.~Barnes, K.~Sun, and S.~Das Sarma, Nat. Comm. \textbf{3}, 997 (2012).


\bibitem{Ivanov11PRA} S.~S.~Ivanov and N.~V.~Vitanov, Phys. Rev. A \textbf{84}, 022319 (2011).

\bibitem{QOptics} S.~S.~Ivanov and N.~V.~Vitanov, Opt. Lett. \textbf{36}, 7 (2011);
 G.~T.~Genov, B.~T.~Torosov, and N.~V.~Vitanov, Phys. Rev. A \textbf{84}, 063413 (2011);
 G.~T.~Genov and N.~V.~Vitanov, Phys. Rev. Lett. \textbf{110}, 133002 (2013).

\bibitem{Torosov11PRA} B.~T.~Torosov and N.~V.~Vitanov, Phys. Rev. A \textbf{83}, 053420 (2011).

\bibitem{Torosov11PRL} B.~T.~Torosov, S.~Gu\'erin, and N.~V.~Vitanov, Phys. Rev. Lett. \textbf{106}, 233001 (2011).

\bibitem{SIvanov13NJP} S.~S.~Ivanov, N.~V.~Vitanov, and N.~V.~Korolkova, New J. Phys. \textbf{15}, 023039 (2013).


\bibitem{Hill2007}
C.~D.~Hill, Phys. Rev. Lett. \textbf{98}, 180501 (2007).

\bibitem{Jones2013pra}
J.~A.~Jones, Phys. Rev. A \textbf{87}, 052317 (2013).

\bibitem{Jones2013pla}
J.~A.~Jones, Phys. Lett. A \textbf{377}, 2860 (2013).

\bibitem{Merrill2014review}
J.~T.~Merrill and K.~R.~Brown, Adv. Chem. Phys. \textbf{154}, 241 (2014)

\bibitem{Cohen2016}
I.~Cohen, A.~Rotem, and A.~Retzker, Phys. Rev. A \textbf{93}, 032340 (2016).

\bibitem{Calderon-Vargas2017}
F.~A.~Calderon-Vargas and J.~P.~Kestner, Phys. Rev. Lett. \textbf{118}, 150502 (2017).

\bibitem{Tomita2010}
Y.~Tomita, J.~T.~Merrill, and K.~R.~Brown, New J. Phys. \textbf{12}, 015002 (2010).


\bibitem{Schraft13PRA} D.~Schraft, T.~Halfmann, G.~T.~Genov, and N.V.~Vitanov, Phys. Rev. A \textbf{88}, 063406 (2013).

\bibitem{Genov2014PRL} G.~T.~Genov, D.~Schraft, T.~Halfmann, and N.~V.~Vitanov, Phys. Rev. Lett. \textbf{113}, 043001 (2014).

\bibitem{Genov2018PRA} G.~T.~Genov, D.~Schraft, and T.~Halfmann, Phys. Rev. A \textbf{98}, 063836 (2018).


\bibitem{RDD_review12Suter} A.~Souza, G.~A.~Alvarez, and D.~Suter, Phil. Trans. R. Soc. A \textbf{370}, 4748-4769 (2012).

\bibitem{CasanovaPRA2015} J.~Casanova, Z.-Y.~Wang, J.~F.~Haase, and M.~B.~Plenio, Phys. Rev. A \textbf{92}, 042304 (2015).

\bibitem{GenovPRL2017} G.~T.~Genov, D.~Schraft, N.~V.~Vitanov, and T.~Halfmann, Phys. Rev. Lett. \textbf{118}, 133202 (2017).

\bibitem{SriarunothaiQST2019} T.~Sriarunothai, S.~W{\"o}lk, G.~S.~Giri, N.~Friis, V.~Dunjko, H.~J.~Briegel, and C.~Wunderlich, Quantum Sci. Technol. \textbf{4}, 015014 (2019).

\bibitem{GenovQST2019} G.~T.~Genov, N.~Aharon, F.~Jelezko, and A.~Retzker, Quantum Sci. Technol. \textbf{4}, 035010 (2019).

\bibitem{ZhouArxuv2019}
H.~Zhou, J.~Choi, S.~Choi, R.~Landig, A.~M.~Douglas, J.~Isoya, F.~Jelezko, S.~Onoda, H.~Sumiya, P.~Cappellaro, H.~S.~Knowles, H.~Park, and M.~D.~Lukin, arXiv:1907.10066 (2019)

\bibitem{Dunning2014PRA} A.~Dunning, R.~Gregory, J.~Bateman, N.~Cooper, M.~Himsworth,
J.~A.~Jones, and T.~Freegarde, Phys. Rev. A \textbf{90}, 033608 (2014).

\bibitem{Vitanov2015PRA} N.~V.~Vitanov, T.~F.~Gloger, P.~Kaufmann, D.~Kaufmann, T.~Collath, M.~Tanveer~Baig, M.~Johanning, and C.~Wunderlich, Phys. Rev. A \textbf{91}, 033406 (2015).

\bibitem{Zanon-Willette2018REPP} T.~Zanon-Willette, R.~Lefevre, R.~Metzdorff, N.~Sillitoe,
S.~Almonacil, M.~Minissale, E.~de~Clercq, A.~V.~Taichenachev, V.~I.~Yudin, and E.~Arimondo, Rep. Prog. Phys. \textbf{81}, 09440 (2018).




\bibitem{Gulde2003}
S.~Gulde, M.~Riebe, G.~P.~T.~Lancaster, C.~Becher, J.~Eschner, H.~H\"{a}ffner, F.~Schmidt-Kaler, I.~L.~Chuang, and R.~Blatt, Nature
\textbf{421}, 48 (2003). 

\bibitem{Schmidt-Kaler2003}
F.~Schmidt-Kaler, H.~H\"{a}ffner, M.~Riebe, S.~Gulde, G.~P.~T.~Lancaster, T.~Deuschle, C.~Becher, C.~F.~Roos, J.~Eschner, and R.~Blatt,
Nature \textbf{422}, 408 (2003). 

\bibitem{Haffner2008} H.~H\"{a}ffner, C.~F.~Roos, and R.~Blatt, Phys. Rep. \textbf{469}, 155 (2008).

\bibitem{Timoney2008} N.~Timoney, V.~Elman, S.~Glaser, C.~Weiss, M.~Johanning, W.~Neuhauser, and C.~Wunderlich, Phys. Rev. A
    \textbf{77}, 052334 (2008).

\bibitem{Monz2009} T.~Monz, K.~Kim, W.~H\"{a}nsel, M.~Riebe, A.~S.~Villar, P.~Schindler, M.~Chwalla, M.~Hennrich, and R.~Blatt, Phys.
    Rev. Lett. \textbf{102}, 040501 (2009).

\bibitem{Shappert2013}
C.~M.~Shappert, J.~T.~Merrill, K.~R.~Brown, J.~M.~Amini, C.~Volin, S.~C.~Doret, H.~Hayden, C.~S.~Pai, and A.~W.~Harter, New J. Phys.
\textbf{15}, 083053 (2013).

\bibitem{Mount2015}
E.~Mount, C.~Kabytayev, S.~Crain, R.~Harper, S.-Y.~Baek, G.~Vrijsen, S.~T.~Flammia, K.~R.~Brown, P.~Maunz, and J.~Kim, Phys. Rev. A
\textbf{92}, 060301(R) (2015).

 \bibitem{Rakreungdet2009}
W.~Rakreungdet, J.~H.~Lee, K.~F.~Lee, B.~E.~Mischuck, E.~Montano, and P.~S.~Jessen, Phys. Rev. A \textbf{79}, 022316 (2009).

\bibitem{Butts2013}
D.~L.~Butts, K.~Kotru, J.~M.~Kinast, A.~M.~Radojevic, B.~P.~Timmons, and R.~E.~Stoner, J. Opt. Soc. Am. B \textbf{30}, 922 (2013).

\bibitem{Dunning2014}
A.~Dunning, R.~Gregory, J.~Bateman, N.~Cooper, M.~Himsworth, J.~A.~Jones, and T.~Freegarde,
Phys. Rev. A \textbf{90}, 033608 (2014).

\bibitem{Berg2015}
P.~Berg, S.~Abend, G.~Tackmann, C.~Schubert, E.~Giese, W.~P.~Schleich, F.~A.~Narducci, W.~Ertmer, and E.~M.~Rasel, Phys. Rev. Lett.
\textbf{114}, 063002 (2015).

\bibitem{Demeter2016}
G.~Demeter, Phys. Rev. A \textbf{93}, 023830 (2016).

\bibitem{Wang2012}
X.~Wang, L.~S.~Bishop, J.~P.~Kestner, E.~Barnes, K.~Sun, and S.~Das~Sarma, Nature Commun. \textbf{3}, 997 (2012).

\bibitem{Kestner2013}
J.~P.~Kestner, X.~Wang, L.~S.~Bishop, E.~Barnes, and S.~Das~Sarma, Phys. Rev. Lett. \textbf{110}, 140502 (2013).

\bibitem{Wang2014}
X.~Wang, L.~S.~Bishop, E.~Barnes, J.~P.~Kestner, and S.~Das~Sarma, Phys. Rev. A \textbf{89}, 022310 (2014).

\bibitem{Zhang2017}
C.~Zhang, R.~E.~Throckmorton, X.-C.~Yang, X.~Wang, E.~Barnes, and S.~Das~Sarma, Phys. Rev. Lett. \textbf{118}, 216802 (2017).

\bibitem{Hickman2013}
G.~T.~Hickman, X.~Wang, J.~P.~Kestner, and S.~Das~Sarma, Phys. Rev. B \textbf{88}, 161303(R) (2013).

\bibitem{Eng2015}
K.~Eng, T.~D.~Ladd, A.~Smith, M.~G.~Borselli, A.~A.~Kiselev, B.~H.~Fong, K.~S.~Holabird, T.~M.~Hazard, B.~Huang, P.~W.~Deelman, I.~Milosavljevic, A.~E.~Schmitz, R.~S.~Ross, M.~F.~Gyure, and A.~T.~Hunter, Sci. Adv. \textbf{1}, e1500214 (2015)

\bibitem{Rong2015}
X.~Rong, J.~Geng, F.~Shi, Y.~Liu, K.~Xu, W.~Ma, F.~Kong, Z.~Jiang, Y.~Wu, and J.~Du, Nature Commun. \textbf{6}, 8748 (2015).

\bibitem{Aiello2013}
C.~D.~Aiello, M.~Hirose, and P.~Cappellaro, Nature Commun. \textbf{4}, 1419 (2013).


\bibitem{SchwartzSciAdv2018}
I.~Schwartz, J.~Scheuer, B.~Tratzmiller, S.~Mueller, Q.~Chen, I.~Dhand, Z.~Wang, C.~Mueller, B.~Naydenov, F.~Jelezko, and M.~B.~Plenio, Sci. Adv. \textbf{4}, eaat8978 (2018).

\bibitem{Ventura2019}
C.~Ventura-Vel\'azquez, B.~J.~\'Avila, E.~Kyoseva, and B.~M.~Rodriguez-Lara, Scientific Rep. \textbf{9}, 4382 (2019).
%



\bibitem{Fleischhauer05RMP} M.~Fleischhauer, A.~Imamoglu, and J.~P.~Marangos, Rev. Mod. Phys. \textbf{77}, 633 (2005).


\bibitem{Heinze13PRL} G.~Heinze, C.~Hubrich, and T.~Halfmann, Phys. Rev. Lett. \textbf{111}, 033601 (2013).

\bibitem{Schraft16PRL} D.~Schraft, M.~Hain, N.~Lorenz, and T.~Halfmann, Phys. Rev. Lett. \textbf{116}, 073602 (2016) 

\bibitem{Mieth14OE} S.~Mieth, A.~Henderson, and T.~Halfmann, Opt. Express \textbf{22}, 011182 (2014) 


\bibitem{CPMG} H.~Y.~Carr and E.~M.~Purcell, Phys. Rev. \textbf{94}, 630 (1954); S.~Meiboom and D.~Gill, Rev. Sci. Instrum. \textbf{29}, 688-691 (1958); U.~Haeberlen, \emph{High Resolution NMR in Solids: Selective Averaging} (Academic, New York, 1976); W.~M.~Witzel and S.~Das~Sarma, Phys. Rev. Lett. \textbf{98}, 077601 (2007).

\bibitem{Mieth12PRA} S.~Mieth, D.~Schraft, T.~Halfmann, and L.~P.~Yatsenko, Phys. Rev. A \textbf{86}, 063404 (2012). 


\bibitem{Pascual-Winter12PRB} M.~F.~Pascual-Winter, R.~C.~Tongning, R.~Lauro, A.~Louchet-Chauvet, T.~Chaneli\`{e}re, and J.-L.~Le Gou\"{e}t, Phys. Rev. B \textbf{86}, 064301 (2012).










\end{thebibliography}
\end{document}